\documentclass{article}

\usepackage[final,nonatbib]{neurips_2024_ml4ps}

\usepackage[utf8]{inputenc} %
\usepackage[T1]{fontenc}    %
\usepackage{hyperref}       %
\usepackage{url}            %
\usepackage{booktabs}       %
\usepackage{amsfonts}       %
\usepackage{nicefrac}       %
\usepackage{microtype}      %
\usepackage{xcolor}         %
\usepackage{graphicx}
\usepackage{tablefootnote}
\usepackage{sidecap}
\usepackage{subcaption}

\newcommand{\mhalo}{{\rm{M}_{\rm{halo}}}}
\newcommand{\mst}{{\rm{M}_{*}}}
\newcommand{\msun}{{\rm{M}_{\odot}}}

\title{Estimating Dark Matter Halo Masses in Simulated Galaxy Clusters with Graph Neural Networks}

\author{%
  Nikhil Garuda \\
  Steward Observatory\\
  University of Arizona\\
  933 N Cherry Ave, Tucson, AZ, 85721 \\
  \texttt{nikhilgaruda@arizona.edu} \\
  \And
  John F. Wu \\
  Space Telescope Science Institute \\
  3700 San Martin Dr, Baltimore, MD 21218 \\
  \texttt{jowu@stsci.edu} 
  \And
  Dylan Nelson \\
  Universität Heidelberg \\
  Zentrum für Astronomie, ITA, \\Albert-Ueberle-Str. 2 \\
  69120 Heidelberg, Germany \\
  \texttt{dnelson@uni-heidelberg.de} \\
  \And
  Annalisa Pillepich \\
  Max-Planck-Institut für Astronomie \\
  Königstuhl 17, 69117 Heidelberg, Germany \\
  \texttt{pillepich@mpia.de} \\
}

\begin{document}

\maketitle

\begin{abstract}
    Galaxies grow and evolve in dark matter halos. Because dark matter is not visible, galaxies' halo masses ($\mhalo$) must be inferred indirectly. We present a graph neural network (GNN) model for predicting $\mhalo$ from stellar mass ($\mst$) in simulated galaxy clusters using data from the IllustrisTNG simulation suite. Unlike traditional machine learning models like random forests, our GNN captures the information-rich substructure of galaxy clusters by using spatial and kinematic relationships between galaxy neighbour. A GNN model trained on the TNG-Cluster dataset and independently tested on the TNG300 simulation achieves superior predictive performance compared to other baseline models we tested. Future work will extend this approach to different simulations and real observational datasets to further validate the GNN model’s ability to generalise.
\end{abstract}

\section{Introduction} \label{sec:intro}
\vspace{-1em}

In the Lambda Cold Dark Matter cosmological model \cite{peeblesTestsCosmologicalModels1984a, bullockSmallScaleChallengesCDM2017}, galaxies form and evolve in dark matter halos. Cosmological simulations demonstrate that galaxies grow in tandem with their dark matter halos according to well-measured and tight scaling relations \cite{wechslerConnectionGalaxiesTheir2018}. This interdependence between stellar mass ($\mst$) and subhalo mass ($\mhalo$) is known as the stellar–halo mass relation (SHMR).

While $\mst$ is observable, $\mhalo$ must often be inferred indirectly via the SHMR due to observational constraints. For example, galaxy clusters---the most massive gravitationally bound objects in the Universe---are dark matter dominated, but their total mass must be measured via gravitational lensing \cite{cloweDirectEmpiricalProof2006,vegettiStrongGravitationalLensing2024}, the Sunyaev-Zel'dovich effect \cite{birkinshawSunyaevZeldovichEffect1999, 2011ApJ...737...61M, 2015ApJS..216...27B}, and/or visible wavelength proxies (e.g., galaxy richness, intracluster light, etc; \cite{2014ApJ...785..104R,2021MNRAS.501.1300S}). However, these methods are unable to fully leverage galaxy substructure within clusters to estimate their dark matter halo masses.

Therefore, we present a graph neural network (GNN) algorithm \cite{scarselliGraphNeuralNetwork2009} for predicting $\mhalo$ for galaxies in simulated cluster environments\footnote{\url{https://github.com/Nikhil0504/halo_masses}}. Compared to primitive machine learning (ML) methods like random forests \cite{agarwalPaintingGalaxiesDark2018}, a GNN can learn the substructure in neighbouring galaxies and thereby improve halo mass predictions. Our results using the GNN demonstrate significant performance gains on the training, validation, and an independent test set.

\section{IllustrisTNG Simulation Data} \label{sec:data}
\vspace{-1em}

The simulation data we use are large-volume, cosmological, gravo-magnetohydrodynamical simulations from the IllustrisTNG simulation suite \cite{nelsonIllustrisTNGSimulationsPublic2019}. We specifically use the TNG-Cluster \cite{nelsonIntroducingTNGClusterSimulation2024a} simulation, a collection of zoom-in simulations centered 352 of the most massive halos (i.e., galaxy clusters), for training and validation. Our dataset is based on the \texttt{SUBFIND} \cite{springelPopulatingClusterGalaxies2001} subhalo catalogs that were obtained from snapshot 99 ($z = 0$), focusing on the high-resolution components of the zoom-in simulation. We adopt cosmological parameters from \cite{planckcollaborationPlanck2015Results2016}, using $H_0 = 67.74$ km s$^{-1}$ Mpc$^{-1}$ for consistency with the IllustrisTNG simulation suite. Additional details about the TNG-Cluster training data are provided in Appendix \ref{sec:appendix-tng-cluster-info}. The distribution of subhalos in TNG-Cluster is shown in Figure \ref{fig:halo-distr-tng-cluster}, and the selection criteria and number of samples are described in Table~\ref{tab:sim-cuts}.

We test our ML models on an independent data set, the Illustris TNG300-1 hydrodynamic simulation (hereafter TNG300; \cite{nelsonIllustrisTNGSimulationsPublic2019}). The TNG-Cluster and TNG300 simulations use the same physics and have comparable resolutions (in the former's zoom-in regions), but the two simulations are otherwise independent. 
When reporting TNG-Cluster cross-validation results TNG300 test set results, we \textit{only} consider galaxies within 10~Mpc of all clusters with $\mhalo >10^{14} \msun$.

\vspace{-1em}
\section{Methods/Experiments} \label{sec:methods-experiments}
\vspace{-1em}

The primary objective of our study is to estimate $\mhalo$ from $\mst$. Building on the work of \cite{larsonPredictingDarkMatter2024}, we train ML models on galaxies and dark matter halos from TNG-Cluster to probe cluster environments. 

\begin{figure}[t]
    \centering
    \includegraphics[width=\linewidth]{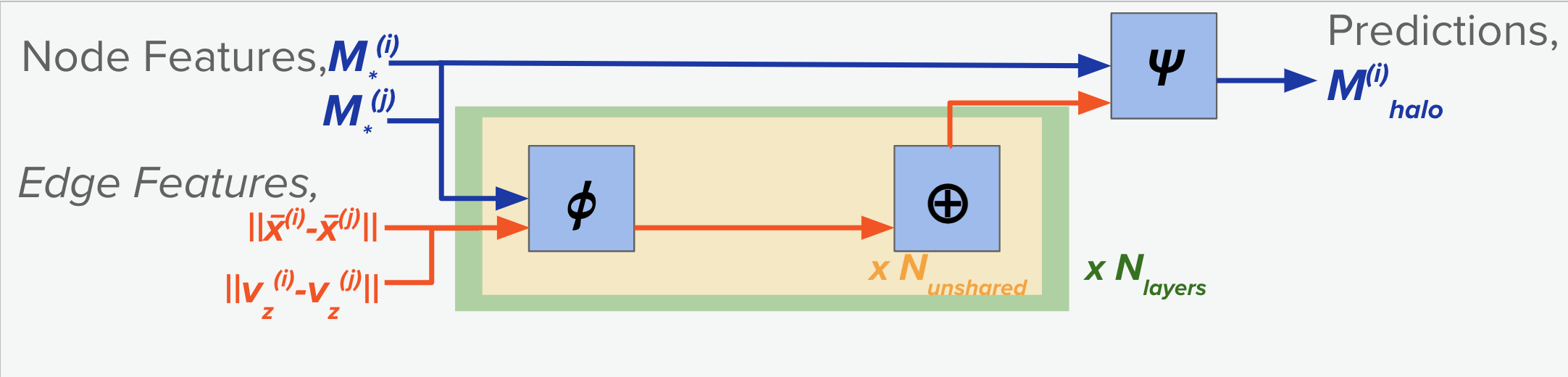}
    \caption{Flow diagram of GNN architecture used for halo mass prediction. The GNN processes node features ($x_i$, $x_j$) and edge features ($\epsilon_{ij}$) through multiple unshared layers, where each layer applies learnable functions, $\phi$, which are implemented as MLPs. These unshared layers operate in parallel across the graph structure. A pooling layer then aggregates ($\bigoplus$) the information from these interactions back into each node. Subsequent repetitions of these GNN layers can give it more representational power. Finally, the output MLP, $\psi$, combines node features and aggregated edge features to predict each node's halo mass.}\vspace{-1em}
    \label{fig:gnn-flow-dig}
\end{figure}

\vspace{-1em}
\subparagraph{Loss Functions and Evaluation Metrics.} Model performance is assessed using several metrics (presented in Table~\ref{tab:metrics}). Simple models are trained to minimise the Mean Squared Error (MSE), while the GNN is optimised using Gaussian negative log-likelihood (combining MSE and log-variance terms, per \cite{jeffreySolvingHighdimensionalParameter2020})\footnote{The negative log-likelihood objective accounts for the intrinsically varying scatter in $\mhalo$.}. Validation and test performance are evaluated with Root Mean Squared Error (RMSE), Mean Absolute Error (MAE), coefficient of determination $R^2$, Normalised Median Absolute Deviation (NMAD), average offset (Bias), and Outlier Fraction (f$_{\rm{outlier}}$).

\vspace{-1em}
\subparagraph{Random Forest Baseline Models.} To establish a benchmark for subsequent comparisons with our GNN model, we use Random Forest (RF) regression \cite{hoRandomDecisionForests1995} as a baseline model due to its capability to handle complex non-linear relationships between features. To further augment the simple RF model, we compute an overdensity parameter ($\Delta_G$), defined as the sum of stellar masses within a specified radius $R_{\rm{max}}$.
The RF models are configured with 100 estimators using \texttt{scikit-learn} \cite{pedregosaScikitlearnMachineLearning2011}, one of which utilises $\mst$, and one which uses both $\mst$ and $\Delta_G$ as features.

\vspace{-1em}
\subparagraph{Graph Neural Networks.} In our GNN model, each node represents a galaxy, with the $\mst$ as the sole node feature. 
We construct edges between galaxy pairs separated by less than 3 Mpc \cite{wuHowGalaxyHaloConnection2024}, connecting neighbouring nodes. 
These connections enable the neural network to learn interactions between the substructure and galaxy properties within the cluster.
We provide two edge features to incorporate both the spatial and kinematic separations of galaxies: the squared Euclidean distance between pairs of galaxy positions, and pairs of relative line-of-sight velocities.

\vspace{-1em}
\subparagraph{GNN Architecture.} Our GNN follows the architecture described in \cite{wuHowGalaxyHaloConnection2024} with 8 unshared layers and 3 sequential layers, as shown in Figure \ref{fig:gnn-flow-dig}. Each layer is composed of a two-layer MLP with 16 hidden channels, SiLU activations \cite{hendrycksGaussianErrorLinear2016}, and 16 outputs. These operate over edges connected to each node, using max pooling to aggregate edge information to each node \footnote{This helps the GNN to effective capture the neighbouring features.}. The node output is concatenated with its initial feature ($\mst$) and passed through a 3-layer MLP. The GNN predicts two quantities \cite{jeffreySolvingHighdimensionalParameter2020}: $\mhalo$ and the expected log variance of $\mhalo$ at a given $\mst$.

\vspace{-1em}
\subparagraph{GNN Optimisation.} We employ the METIS algorithm to partition the training set into 48 parts (see \texttt{ClusterLoader} class in PyTorch Geometric \cite{chiangClusterGCNEfficientAlgorithm2019, feyFastGraphRepresentation2019}), allowing us to handle large graph datasets efficiently. The model is trained with the AdamW optimiser \cite{loshchilovDecoupledWeightDecay2019} at an initial learning rate of $10^{-2}$ and weight decay of $10^{-4}$. A scheduler reduces the learning rate by 0.2 if validation loss stagnates by $10^{-3}$ for 15 epochs. Early stopping occurs after 35 epochs of no improvement, with a maximum of 300 epochs. On an Nvidia A6000 GPU, training takes 20 minutes and inference takes under 1 second.

\vspace{-1em}
\section{Results}
\begin{figure*}[t]
    \centering
    \begin{minipage}{0.5\linewidth}
        \centering
        \includegraphics[width=\linewidth]{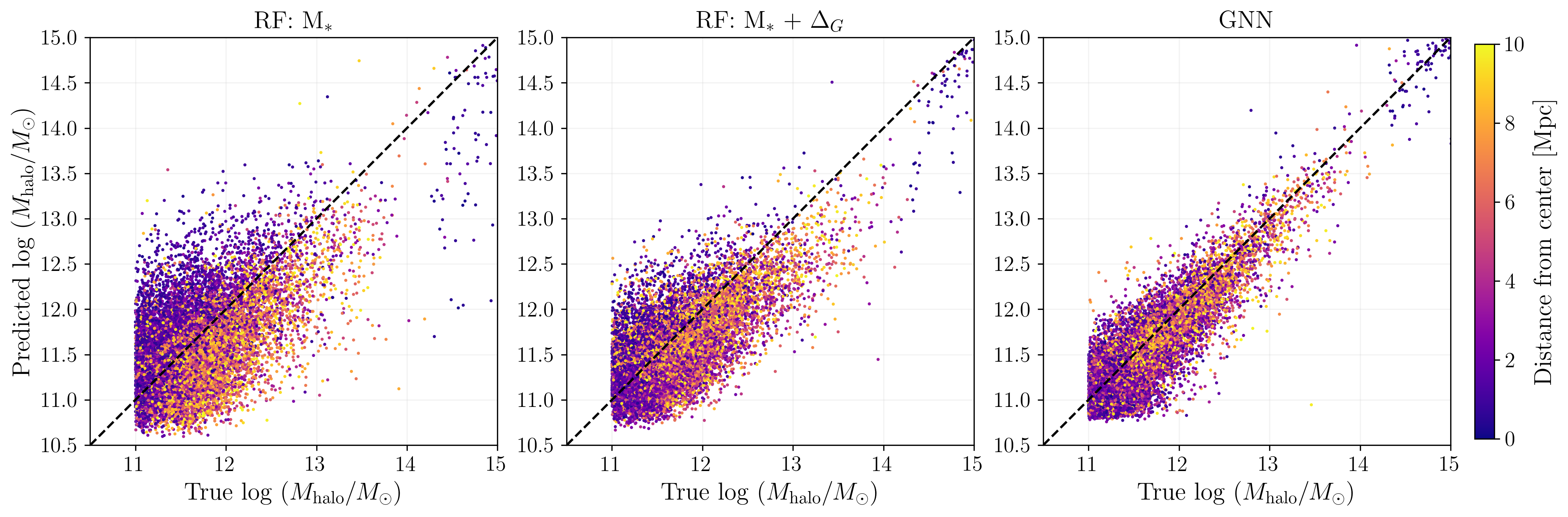}
        \subcaption{Predicted versus true $\mhalo$ for the TNG-Cluster validation set, coloured by distance from cluster center.}
        \label{fig:true-vs-pred}
    \end{minipage}
    \hfill
    \begin{minipage}{0.49\linewidth}
        \centering
        \includegraphics[width=\linewidth]{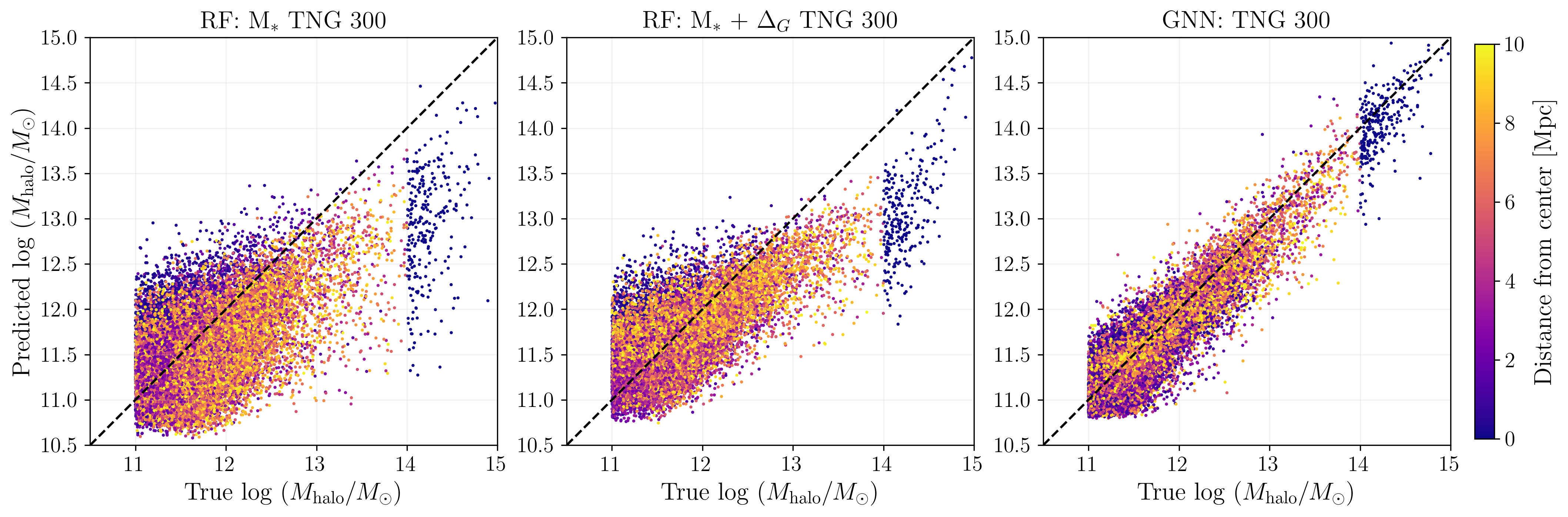}
        \subcaption{Predicted versus true $\mhalo$ for the TNG300 test set, coloured by distance from cluster center.}
        \label{fig:tng300-scatterplot}
    \end{minipage}
    \caption{For each subfigure, we show results for the RF with only $\mst$ as a feature (\textit{left}), the RF with $\mst$ and overdensity parameter $\Delta_G$ (\textit{center}), and the GNN with $\mst$ and graph connectivity (\textit{right}).}
    \label{fig:combined}
\end{figure*}

\vspace{-1em}
Table \ref{tab:metrics} compares model performance for predicting $\mhalo$ from galaxies residing in clusters for the validation and test datasets. We additionally show the scatter of $\mhalo$ in the first row, which represents the most naive ``prediction'' of the sample mean. Below, we present the results for the baseline models and GNN model. We display scatter plots of the true versus predicted masses for the TNG-Cluster cross-validation data set in Figure \ref{fig:true-vs-pred} and TNG300 test set in Figure \ref{fig:tng300-scatterplot}.

\begin{table}[t] 
\caption{Validation and test set performance for all models. The best metrics are underlined.}
\centering
\resizebox{\textwidth}{!}{%
\begin{tabular}{@{}lllllll@{}}
\toprule
Model                                              & RMSE                                       & MAE                                        & $R^2$                                      & Bias                                        & f$_{\rm{outlier}}$                         & NMAD                                       \\ \midrule
TNG-Cluster cross-validation                       &                                            &                                            &                                            &                                             &                                            &                                            \\
\hspace{1em}(Always predict mean) & $0.542$                                      & $0.396$                                      & $0$                                          & $0$                                           & $0.019$                                      & $0.479$                                      \\
\hspace{1em}RF: $\mst$            & $0.489_{\pm 0.002}$                        & $0.382_{\pm 0.003}$                        & $0.186_{\pm 0.011}$                        & \underline{$-0.067_{\pm 0.006}$}                        & \underline{$0.008_{\pm 0.001}$} & $0.463_{\pm 0.005}$                                   \\
\hspace{1em}RF: $\mst + \Delta_G$ & $0.385_{\pm 0.002}$                        & $0.301_{\pm 0.002}$                        & $0.490_{\pm 0.007}$                        & $-0.124_{\pm 0.004}$                        & $0.008_{\pm 0.000}$                        & $0.367_{\pm 0.002}$                        \\
\hspace{1em}GNN                   & \underline{$0.273_{\pm 0.010}$} & \underline{$0.209_{\pm 0.009}$} & \underline{$0.745_{\pm 0.019}$} & $-0.085_{\pm0.027}$ & $0.013_{\pm 0.002}$                        & \underline{$0.246_{\pm 0.013}$} \\ \midrule
TNG-300 test set                                   &                                            &                                            &                                            &                                             &                                            &                                            \\
\hspace{1em}(Always predict mean) & $0.466$                                      & $0.351$                                      & $0$                                          & $0$                                           & $0.021$                                      & $0.422$                                      \\
\hspace{1em}RF                    & $0.468_{\pm 0.009}$                        & $0.365_{\pm 0.014}$                        & $0.199_{\pm 0.030}$                        & $-0.200_{\pm 0.017}$                        & \underline{$0.009_{\pm 0.003}$}                        & $0.456_{\pm 0.033}$                        \\
\hspace{1em}RF: $\mst + \Delta_G$ & $0.344_{\pm 0.003}$                        & $0.256_{\pm 0.003}$                        & $0.567_{\pm 0.007}$                        & $-0.048_{\pm 0.001}$                        & $0.022_{\pm 0.001}$                        & $0.293_{\pm 0.006}$                        \\
\hspace{1em}GNN                   & \underline{$0.242_{\pm 0.013}$}                        & \underline{$0.184_{\pm 0.010}$}                        & \underline{$0.785_{\pm 0.023}$}                        & \underline{$-0.039_{\pm 0.034}$}                        & $0.014_{\pm 0.002}$                        & \underline{$0.217_{\pm 0.014}$}                        \\ \bottomrule
\end{tabular}
}
\label{tab:metrics}

\footnotesize{Note: The intrinsic scatter in $\mhalo$ ranges from 0.42 (at $\log(\mhalo)=11 ~\msun$) to 0.33 (at $\log(\mhalo)=13 ~\msun$) dex in TNG-Cluster and 0.48 dex to 0.19 dex in TNG 300 respectively.}
\end{table}

The simplest RF model exhibits high error and very low predictive power.\footnote{In fact, for the TNG300 test set, the simplest RF model produces even higher error than the scatter inherent to the data. We ascribe this to the RF model's significant negative bias (i.e., systematic underprediction).} When we augment the RF model with $\Delta_G$, the performance improves, demonstrating that galaxy environments contain vital information for the SHMR. Nonetheless, the RF models systematically underpredict $\mhalo$ for the highest-mass galaxies and yield high error. 

GNNs greatly outperform RF models, as indicated by the right-most panels of Figures~\ref{fig:true-vs-pred} and \ref{fig:tng300-scatterplot}. 
Running the same experiments using XGBoost (which is more prone to overfitting), we find a significant improvement over RF but not enough to surpass GNNs.
We find that the GNN performance on the training and validation sets translates to accurate predictions on the independent test set. For nearly all metrics in Table~\ref{tab:metrics}, the GNN outperforms the RF models for cross-validation and test sets.

\vspace{-1em}
\section{Discussion} 
\label{sec:discussions}
\vspace{-1em}

\subsection{Model performance as a function of local environment}
\vspace{-0.5em}

\begin{figure}[t]
    \centering
    \includegraphics[width=0.5\linewidth]{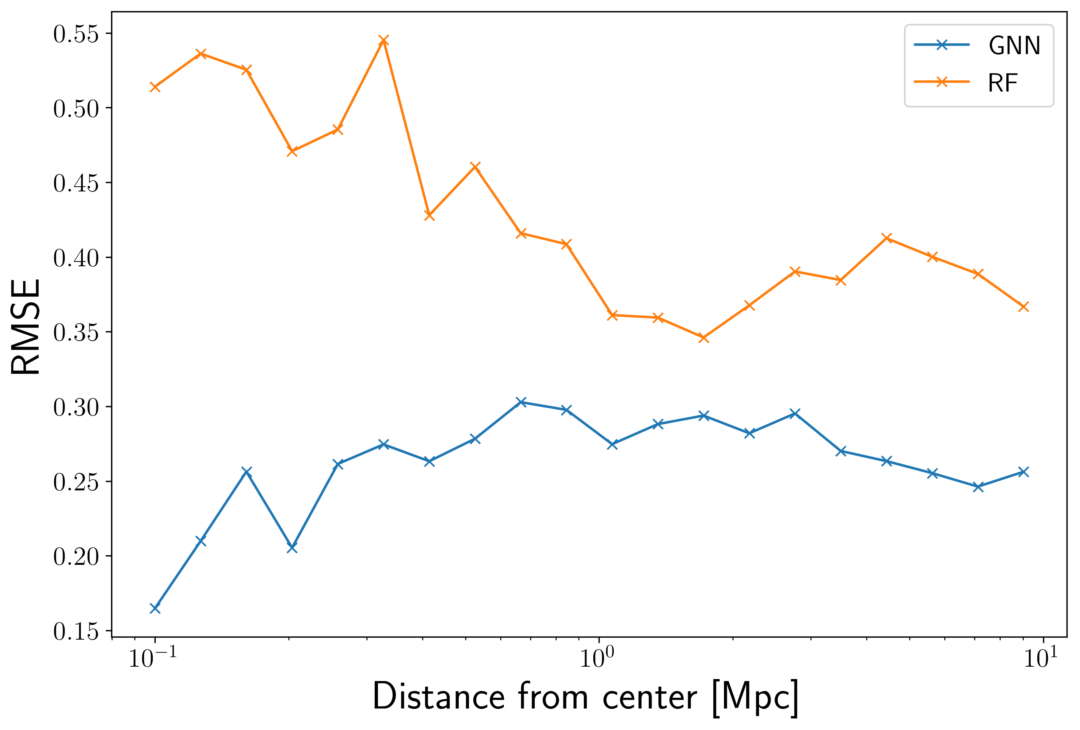}
    \caption{Validation set RMSE as a function of distance from cluster center. Results shown for the GNN (blue) and RF with $\mst$ and $\Delta_G$ (orange).\vspace{-1em}}
    \label{fig:rmse-dist-from-center}
\end{figure}

In Figure \ref{fig:rmse-dist-from-center}, we show the cross-validation RMSE as a function of distance from the cluster center for the GNN and RF ($\mst$ and $\Delta_G$) models; the GNN significantly outperforms the RF across all distance bins. Notably, the RF model performance suffers for galaxies closer to the center of the cluster. One potential explanation for this discrepancy is that the RF does not account for the dense cluster environment, where interactions such as tidal stripping can lead to significant loss of $\mhalo$.\footnote{Due to line-of-sight effects, not all galaxies at small \textit{projected} distances experience significant tidal stripping.} In contrast, the GNN model outperforms the RF due by leveraging information from galaxy pairwise distances and line-of-sight velocities.

\subsection{Comparison against previous work}
\vspace{-0.5em}
Previous studies have used ML to estimate galaxy properties from dark matter halos \cite{kamdarMachineLearningCosmological2016,agarwalPaintingGalaxiesDark2018}, i.e. the inverse of the problem we tackle. Some works employ feature importance from decision tree-based methods \cite{2022MNRAS.509.5046L,wuHowGalaxyHaloConnection2024}, while others use reinforcement learning to connect halo properties to galaxies \cite{mosterGalaxyNetConnectingGalaxies2021}. Convolutional neural networks (CNNs) and GNNs have also been used to predict galaxy stellar masses from simulated halos \cite{2023MNRAS.526.2812C,wuLearningGalaxyenvironmentConnection2023,wuHowGalaxyHaloConnection2024}. 

Several works have used ML methods to predict cluster halo masses from observable parameters such as X-ray brightness and Sunyaev-Zel'dovich decrements \cite{2019ApJ...876...82N,2023MNRAS.524.3289H}.
\cite{yanGalaxyClusterMass2020} compare how different cluster observables fare when pixelised as inputs to a CNN.
\cite{larsonPredictingDarkMatter2024} use GNNs to predict $\mhalo$ directly from galaxy point clouds, but their training dataset (the much smaller TNG50 simulation) does not contain many rare galaxy clusters. 
Our work is the first to train and test GNNs for predicting halo masses in the extremely overdense regime of galaxy clusters.

\vspace{-1em}
\section{Conclusions, Limitations and Future Work} \label{sec:conclusions-limits-future}
\vspace{-1em}
In this work, we predict $\mhalo$ for simulated galaxies using their stellar masses, $2D$ projected positions, and line-of-sight velocities (i.e., $x, y, v_z$) with the TNG-Cluster simulation for training and TNG300 for testing. We evaluated both Random Forest (RF) models and Graph Neural Networks (GNNs). The key findings are:
\begin{enumerate}
    \item The GNN model significantly outperforms RF model, even when the latter is provided $\Delta_G$ as a parameter. This suggests that GNNs capture the underlying spatial relationships and substructures within clusters, as shown in Table~\ref{tab:metrics} and Figures~\ref{fig:combined} and \ref{fig:rmse-dist-from-center}.
    \item The GNN maintains its predictive power when tested on the independent TNG300 dataset, demonstrating that the model generalises across the IllustrisTNG simulation suite. 
\end{enumerate}

Despite our promising results, models trained on one simulation may face challenges when applied to other simulations or real observational data.
Machine learning models are often susceptible to domain shift, where their performance degrades when applied to datasets that differ from their training data \cite{sunReturnFrustratinglyEasy2016, kouwIntroductionDomainAdaptation2019}. In our case, the comparable performance between the TNG-Cluster cross-validation and TNG300 test datasets suggests that the GNN model may be robust to domain shift within the IllustrisTNG suite. 
This robustness could be attributed to the GNN's ability to learn generalizable symbolic relationships \cite{cranmerDiscoveringSymbolicModels2020}.
Further tests using other simulation physics or with observed datasets (e.g., galaxies at other redshifts) are needed before we can conclude that this method is fully generalizable.

In future work, we will account for observational effects like contaminating galaxies in projection, missing data, and photometric redshift uncertainties, as well as broader concerns about domain shift in ML (see e.g. \cite{2021MNRAS.506..677C}). 
Aside from additional validation on other cosmological simulations \cite{schayeFLAMINGOProjectCosmological2023}, we will test on observational data using published $\mhalo$ estimates for well-known galaxy clusters (e.g., \cite{2017ApJ...837...97L, weaverUNCOVERSurveyFirstlook2024a}). With upcoming telescopes like the Roman Space Telescope \cite{spergel2015widefieldinfrarredsurveytelescopeastrophysics} and Rubin Observatory \cite{lsstdarkenergysciencecollaborationLargeSynopticSurvey2012}, we will be able to study GNN applications to large galaxy cluster samples in the wide-field domain.

\bibliographystyle{plain}

\clearpage

\appendix

\section{TNG-Cluster Additional Details} \label{sec:appendix-tng-cluster-info}
\begin{figure}[t]
    \centering
    \includegraphics[width=\linewidth]{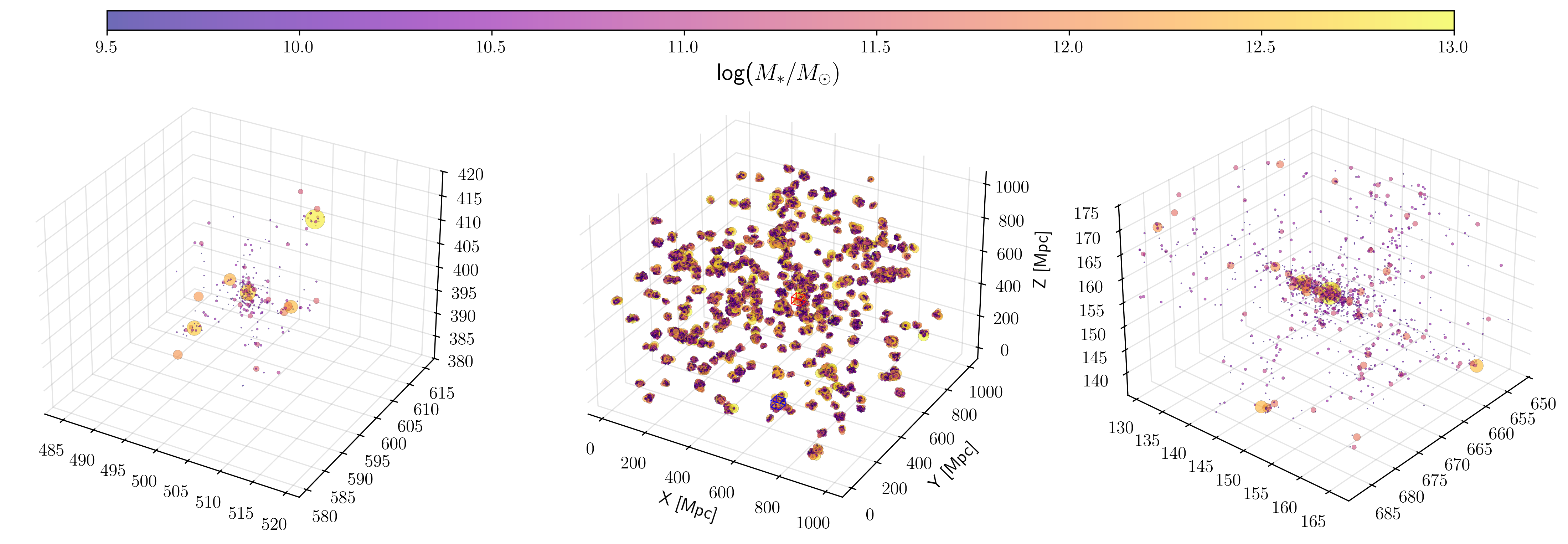}
    \caption{Spatial distribution of halos within the TNG-Cluster simulation. The middle panel shows the full simulation, and the left and right panels highlight two example galaxy clusters. The boundaries of these clusters are marked as blue and red boxes in the middle panel.}
    \label{fig:halo-distr-tng-cluster}
\end{figure}

Galaxies in the TNG-Cluster training data are shown in Figure~\ref{fig:halo-distr-tng-cluster}. To mimic astronomical observations of galaxies, we project the galaxy clusters along the $z$ axis, which is chosen to be the line of sight. This procedure bridges the gap between simulation data and spectroscopic observations, which typically capture two spatial dimensions ($x, y$) and line-of-sight velocities ($v_z$). We also apply quality cuts to the simulation in Table \ref{tab:sim-cuts} to ensure a complete sample of massive, well-resolved galaxies.

\begin{table}[t]
\caption{Summary of cuts applied to the TNG-Cluster data. Here, N$_*$ refers to number of stellar particles, $\msun$ refers to solar mass, R$_{200}$ refers to the virial radius of the halo.}
\centering
\small
\resizebox{\textwidth}{!}{%
\begin{tabular}{@{}lr@{}}
\toprule
Sample       & Number of Subhalos \\ \midrule
Full TNG-Cluster catalog    & 10,378,451           \\
--- within mass cuts - N$_* > 50$; log($\mst / \msun$) $> 9.5$; log($\mhalo / \msun$) $> 10.5$ & 154,120             \\
--- within $< 10\times$ R$_{200}$ of the cluster halo  & 127,165             \\ \midrule 
Selection Criteria - log($\mhalo / \msun$) $> 11$; within 10 Mpc of the cluster halo & \\
TNG-Cluster cross-validation & 60,756 \\
TNG300 Test Set & 34,689 \\ \bottomrule
\end{tabular}
}
\label{tab:sim-cuts}
\end{table}

We split the TNG-Cluster data into training and validation sets by implementing a $k$-fold cross-validation strategy based on cluster IDs rather than traditional random splits. This method isolates subhalos according to their cluster IDs while ensuring that all subhalos from a single cluster remain within the same fold. One potential caveat of this method is that we do not include the contaminating structure along the line-of-sight from other clusters which might be in a different $k$-fold.

\end{document}